\documentclass[twocolumn,preprintnumbers,amsmath,amssymb,prb]{revtex4}

\usepackage{graphicx}
\usepackage{dcolumn}
\usepackage{bm}
\usepackage{amssymb}
\usepackage{epstopdf}
\usepackage{color}

\usepackage{subfiles}

\begin{document}

\title{Floquet engineering of the Lifshitz phase transition in the  Hubbard model}

\author{I. V. Iorsh$^{1,2,3}$}\email{i.iorsh@metalab.ifmo.ru}
\author{D. D. Sedov$^{2}$}
\author{S. A. Kolodny$^{2,3}$}
\author{R. E. Sinitskiy$^3$}
\author{O. V. Kibis$^3$}
\affiliation{$^1$Abrikosov Center for Theoretical Physics, MIPT~University, Dolgoprudny 141701, Russia}
\affiliation{$^2$Department of Physics and Engineering, ITMO~ University, Saint-Petersburg 197101, Russia}
\affiliation{$^3$Department of Applied and Theoretical Physics, Novosibirsk~State~Technical~University,
Karl~Marx~Avenue~20,~Novosibirsk~630073,~Russia}

\begin{abstract}
Within the Floquet theory of periodically driven quantum systems, we demonstrate that an off-resonant high-frequency electromagnetic field can induce the Lifshitz phase transition in periodical structures described by the one-dimensional repulsive Hubbard model with the nearest and next-nearest-neighbor hopping. The transition changes the topology of electron energy spectrum at the Fermi level, transforming it from the two Fermi points to the four Fermi points, which facilitates the emergence of the superconducting fluctuations in the structure. Possible manifestations of the effect and conditions of its experimental observability are discussed.
\end{abstract}
\maketitle

\section{Introduction}
Controlling electronic properties by an off-resonant high-frequency electromagnetic field, which is based on the Floquet theory of periodically driven quantum system (``Floquet engineering''), has been thoroughly explored in last decades and by now is an established research area ~\cite{casas2001floquet,eckardt2015high,bukov2015universal,goldman2014periodically,basov2017towards,oka2019floquet} with numerous predicted and experimentally observed phenomena in atomic systems~\cite{sorensen2005fractional,eckardt2005analog,bermudez2011synthetic,kolovsky2011creating,hauke2012non,Kibis_2022,Delone_book}, quantum circuits~\cite{son2009floquet,kyriienko2018floquet,deng2015observation, PhysRevLett.130.023601}, solid state systems~\cite{oka2009photovoltaic,kibis2010metal,mciver2020light,kibis2017all} and nanostructures~\cite{kibis2011dissipationless,yin2011observation,glazov2014high,wang2013observation,park2022steady,Iorsh_2022,Iurov_2022}.
Since the off-resonant field cannot be absorbed by electrons, it only dresses them, modifying all electronic properties. Such a dressing leads both to the renormalization of the existing terms in the electron Hamiltonian and to the emergence of new terms (e.g. the spin-orbit coupling~\cite{ostermann2019cavity}), what drastically changes band structure and electronic transport.
Particularly, the electromagnetic dressing results in the substantial modification of the electron interactions, inducing the electron states bound at repulsive potentials~\cite{KovalevKibis2020}, the electron pairing in systems containing charge carriers with different effective masses~\cite{Kibis_2019,kibis2021optically} and the new interactions (such as pair-hoppings and density dependent tunneling) in the electronic systems with the non-parabolic dispersion (e.g., within the simplest one-dimensional single-band Hubbard model)~\cite{itin2015effective}. The interplay of modified electron dispersion and electron interactions leads to the emergence of the many-body phase transitions in the driven systems, including such correlated phases as the Kitaev spin liquids~\cite{kumar2022floquet,banerjee2023emergent,sun2023engineering} and the superconducting phases arisen from the modification of interaction potential~\cite{PhysRevResearch.3.023039} or the suppression of competing correlated phases (e.g., charge density waves~\cite{PhysRevB.101.224506, PhysRevB.91.184506}).

In the present article, we demonstrate theoretically that a dressing electromagnetic field induces changing the topology of the Fermi surface (the Lifshitz phase transition) for a periodical structure described by the one-dimensional single-band Hubbard model with the nearest and next-nearest neighbor hopping. As a consequence of the phase transition, the optically induced superconductivity can appear. The article is organized as follows. In Sec.~II, the effective stationary Hamiltonian of the considered structure is derived within the conventional Floquet theory of periodically driven quantum systems. In Sec.~III, the optically induced Lifshitz transition and the associated superconductivity are discussed. The last two sections contain conclusion and acknowledgements.

\section{Model}
\begin{figure}[t]
\includegraphics[width=1.\columnwidth]{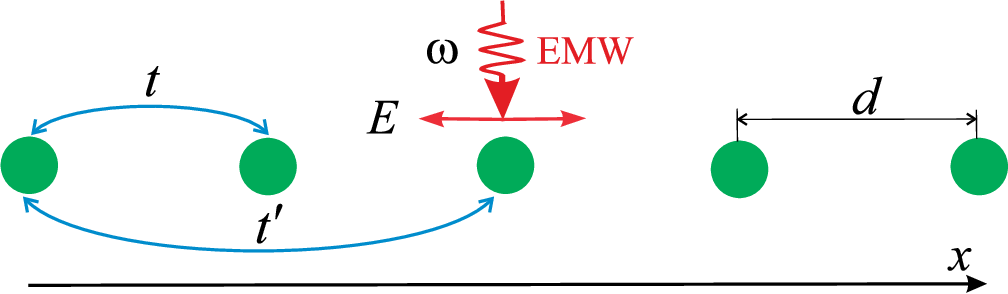}
\caption{Sketch of the system under consideration: A one-dimensional periodic structure consisting of identical unit cells marked by circles and irradiated by a linearly polarized electromagnetic wave (EMW) with the electric field amplitude $E$ and the frequency $\omega$, where $t$ and $t'$ are the nearest and next-nearest neighbor hopping coefficients, respectively, describing the electron transitions between the unit cells, and $d$ is the period of the structure.}
\label{fig:1}
\end{figure}
Let us consider a one-dimensional {infinite} periodical structure irradiated by a linearly polarized off-resonant electromagnetic field within the conventional Hubbard model with the next and next-nearest neighbor hoppings (see Fig.~1). The electron Hamiltonian of the system reads
\begin{eqnarray}\label{H}
\hat{\cal H}(\tau)&=&\sum_{k,\sigma}\varepsilon(k-eA(\tau)/\hbar)\hat{c}_{k\sigma}^{\dagger}\hat{c}_{k\sigma}\nonumber\\
&+&
\sum_{k,k',q} V_0 \hat{c}_{k-q,\uparrow}^{\dagger}\hat{c}_{k'+q,\downarrow}^{\dagger}\hat{c}_{k'\downarrow}\hat{c}_{k\uparrow},
\end{eqnarray}
where $\hat{c}_{k\sigma}^{\dagger}(\hat{c}_{k\sigma})$ are the creation (annihilation) operators,
\begin{equation}\label{ek}
\varepsilon(k)=-t\cos(kd)-t'\cos(2kd)
\end{equation}
is the electron energy spectrum in the field's absence, $k$ is the {continuous} electron wave vector {defined within the first Brillouin zone}, $A(\tau)=E\sin(\omega \tau)/\omega$ is the vector potential of the field, $\tau$ is the time, $E$ is the electric field amplitude, $\omega$ is the field frequency assumed to be far from all resonant frequencies of the electron system, $\sigma=\uparrow,\downarrow$ is the spin quantum number, $d$ is the period of the structure, $t$ and $t'$ are the nearest and next-nearest neighbor hopping coefficients, $V_0$ is the potential of the on-site Coulomb repulsion of electrons, and $q$ is the wavevector transferred in the process of electron-electron interaction. {It should be noted that we consider an off-resonant field, assuming the field frequency to be far from resonant frequencies of the electronic system. As a consequence, the field absorption is suppressed and, correspondingly, the heating of the system can be neglected.}

Within the conventional Floquet theory of periodically driven quantum systems, one can introduce the unitary transformation $\hat{\cal U}(\tau)=e^{i\hat{S}(\tau)}$, which transforms the periodically time-dependent Hamiltonian (\ref{H}) into the effective stationary Hamiltonian
\begin{equation}\label{H0}
\hat{\cal H}_{\mathrm{eff}}=\hat{\cal U}(\tau)^\dagger\hat{\cal H}\hat{\cal U}(\tau) -
i\hbar\hat{\cal U}^\dagger(\tau)\partial_\tau
\hat{\cal U}(\tau).
\end{equation}
There is the regular method to find the transformation operator $\hat{S}(\tau)$ in the case of high-frequency field which satisfies the condition $\hat{\cal H}_n/\hbar\omega\ll1$, where
$\hat{\cal H}_n$ are the components of the Fourier expansion of the time-dependent Hamiltonian \eqref{H},
\begin{equation}\label{Hn}
\hat{\cal H}(\tau)=\sum_{n=-\infty}^{\infty}\hat{\cal H}_ne^{in\omega\tau}.
\end{equation}
Namely, both the operator $\hat{S}(t)$ and the stationary Hamiltonian (\ref{H0}) can be found as an $1/\omega$-expansion (the Floquet-Magnus expansion)~\cite{casas2001floquet,eckardt2015high,bukov2015universal,goldman2014periodically}, which leads to the effective stationary Hamiltonian
\begin{equation}\label{Hef}
\hat{\cal H}_{\mathrm{eff}}=\hat{\cal H}_{0}+\sum_{n=1}^\infty\frac{[\hat{\cal H}_{n},\hat{\cal H}_{-n}]}{n\hbar\omega}+{\it o}\left(\frac{\hat{\cal H}_{n}}{\hbar\omega}\right),
\end{equation}
In the following, we will restrict the analysis by the high-frequency limit,
\begin{equation}\label{c}
t/\hbar\omega\ll1.
\end{equation}
Under this condition, one can take into account only the main term of the expansion (\ref{Hef}), which reads
\begin{equation}\label{H00}
\hat{\cal
H}_0=\sum_{k,\sigma}\tilde{\varepsilon}(k)\hat{c}_{k\sigma}^{\dagger}\hat{c}_{k\sigma}
+\sum_{k,k',q} V_0 \hat{c}_{k-q,\uparrow}^{\dagger}\hat{c}_{k'+q,\downarrow}^{\dagger}\hat{c}_{k'\downarrow}\hat{c}_{k\uparrow},
\end{equation}
where
\begin{equation}\label{ekm}
\tilde{\varepsilon}(k)=-\tilde{t}\cos(kd)-\tilde{t}'\cos(2kd)
\end{equation}
is the electron energy spectrum modified by the field,
$\tilde{t}=tJ_0(\eta)$ and $\tilde{t}'=t'J_0(2\eta)$ are the hopping coefficients modified by the field (dressed hopping coefficients), $J_0(\eta)$ is the Bessel function of the first kind, and $\eta=eEd/\hbar\omega$ is the dimensionless parameter describing the strength of electron-field interaction in the considered system. The effective Hamiltonian \eqref{H00} derived under the condition \eqref{c} has the clear physical meaning. Namely, the condition \eqref{c} can be rewritten as $\tau_0/T\gg1$, where $\tau_0\sim h/t$ is the characteristic lifetime of an electron in an unit cell of the periodic structure pictured in Fig.~1, and $T=2\pi/\omega$ is the field period. Evidently, if the lifetime $\tau_0$ much exceeds the field period $T$, the electron ``feels'' only the time-averaged Hamiltonian \eqref{H}, which is described by Eq.~\eqref{H00}. It follows from the Hamiltonian~\eqref{H00}, particularly, that the high-frequency field modifies only the hopping coefficients, whereas the interaction part of the Hamiltonian remains the same in the main order of the Floquet-Magnus expansion. As to neglected terms of the Hamiltonian \eqref{Hef}, they contribute with the smallness $\sim(t/\hbar\omega)^2$ and can be omitted under the condition \eqref{c} (see Appendix for more details).

\section{Results and discussion}

\begin{figure}[h!]
\centering\includegraphics[width=1.0\columnwidth]{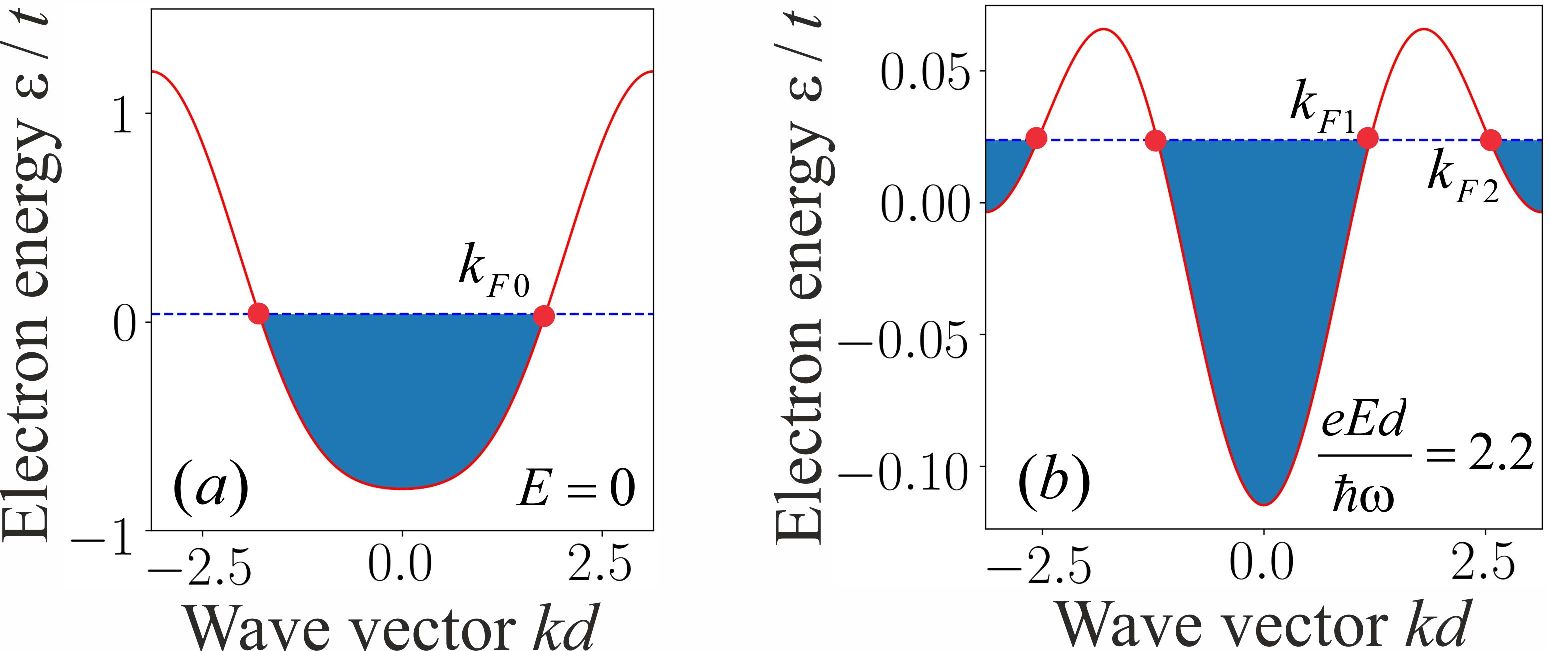}
\caption{Electron energy spectrum near the Fermi level (dashed horizontal line) for the half-filled  system with $t'=0.2t$: (a) in the absence of irradiation; (b) under irradiation with $\eta=eEd/\hbar\omega=2.2$.} \label{fig:2}
\end{figure}
The structure of the electron energy spectrum \eqref{ekm} at the Fermi level depends on the ratio of the dressed hopping coefficients,
\begin{equation}\label{rat}
\frac{\tilde{t}'}{\tilde{t}}=\frac{{t}'}{{t}}\frac{J_0(2\eta)}{J_0(\eta)}.
\end{equation}
Normally, the hopping coefficient $t$ much exceeds the coefficient $t'$. As a consequence, the Fermi surface in the field's absence consists of the two Fermi points with the wave vectors $\pm k_{F0}$ (see Fig.~2b). It follows from Eq.~\eqref{rat} that the dressing field changes the hopping coefficients in the broad range. Particularly, the ratio ~\eqref{rat} can be very large near $\eta=\eta_0$, where $\eta_0\approx2.409$ is the first root of the Bessel function, $J_0(\eta_0)=0$. Therefore, the field crucially changes the Fermi surface, inducing the four Fermi-points with the wave vectors $\pm k_{F1}$ and $\pm k_{F2}$ (see Fig.~2b). As a consequence, the Lifshitz transition with the phase diagram plotted in Fig.~3 appears. It should be noted that the Fermi surface has four points even at very low filling factors near the root $\eta=\eta_0$.
\begin{figure}[h!]
\centering\includegraphics[width=1.0\columnwidth]{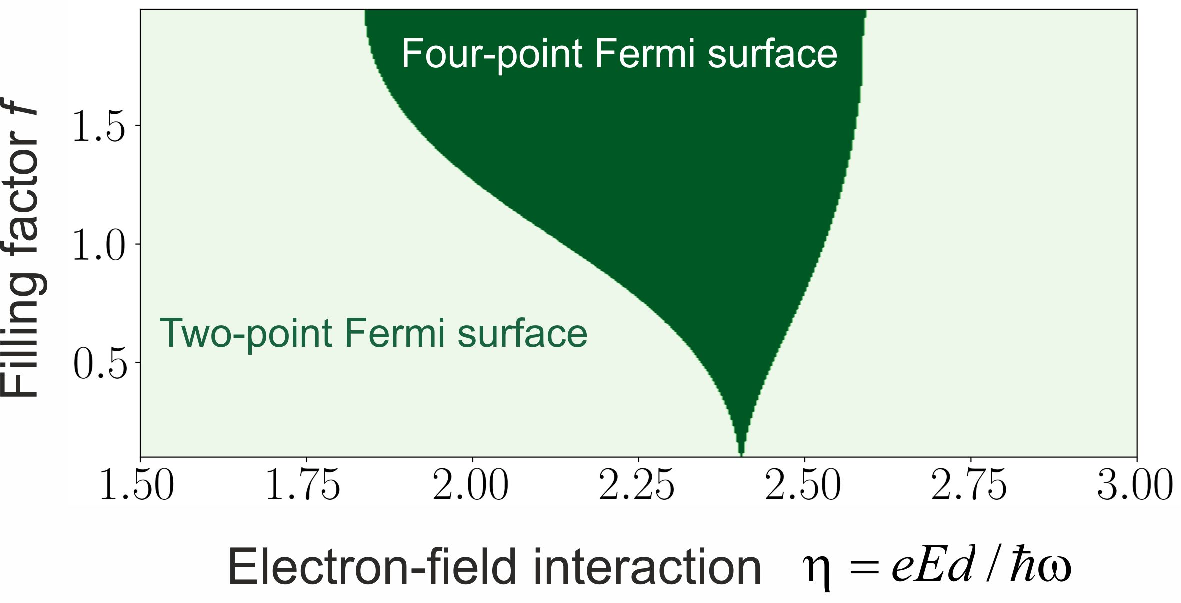}
\caption{Phase diagram of the Lifshitz transition from the two-point Fermi surface to the four-point Fermi surface at $t/t'=0.2$, where the filling factor $f\in[0,\,2]$ is the number of electrons per unit cell.} \label{fig:3}
\end{figure}

To describe the effect of the complex structure of the Fermi surface on the emergence of superconducting instability, one has account for the dynamical screening of the repulsion potential $V_0$ by the electron-hole fluctuations in the vicinity of the Fermi surface. Within the conventional random phase approximation (RPA), the screened potential of electron-electron interaction reads~\cite{mahan2000many}
\begin{align}
V(q)=\frac{V_0}{1+V_0\Pi_q},\quad \Pi_q = \sum_k \frac{f_0(\tilde{\varepsilon}_k)-f_0(\tilde{\varepsilon}_{k+q})}{\tilde{\varepsilon}_{k+q}-\tilde{\varepsilon}_{k}},
\end{align}
where $\Pi_q$ is the static polarizability of the electrons, and $f_0(\varepsilon)$ is the Fermi-Dirac distribution function. In the case of complex Fermi contour, polarizability $\Pi_q$ has peaks at the wave vectors $\pm2k_{F1}$, $\pm2k_{F2}$ and $\pm(k_{F1}-k_{F2})$. This leads to the strong non-monotonic dependence of the screened potential $V(q)$ plotted in Fig.~4a as a function of the wavevector $q$ transferred in the interaction process.
\begin{figure}[h!]
\centering\includegraphics[width=1.0\columnwidth]{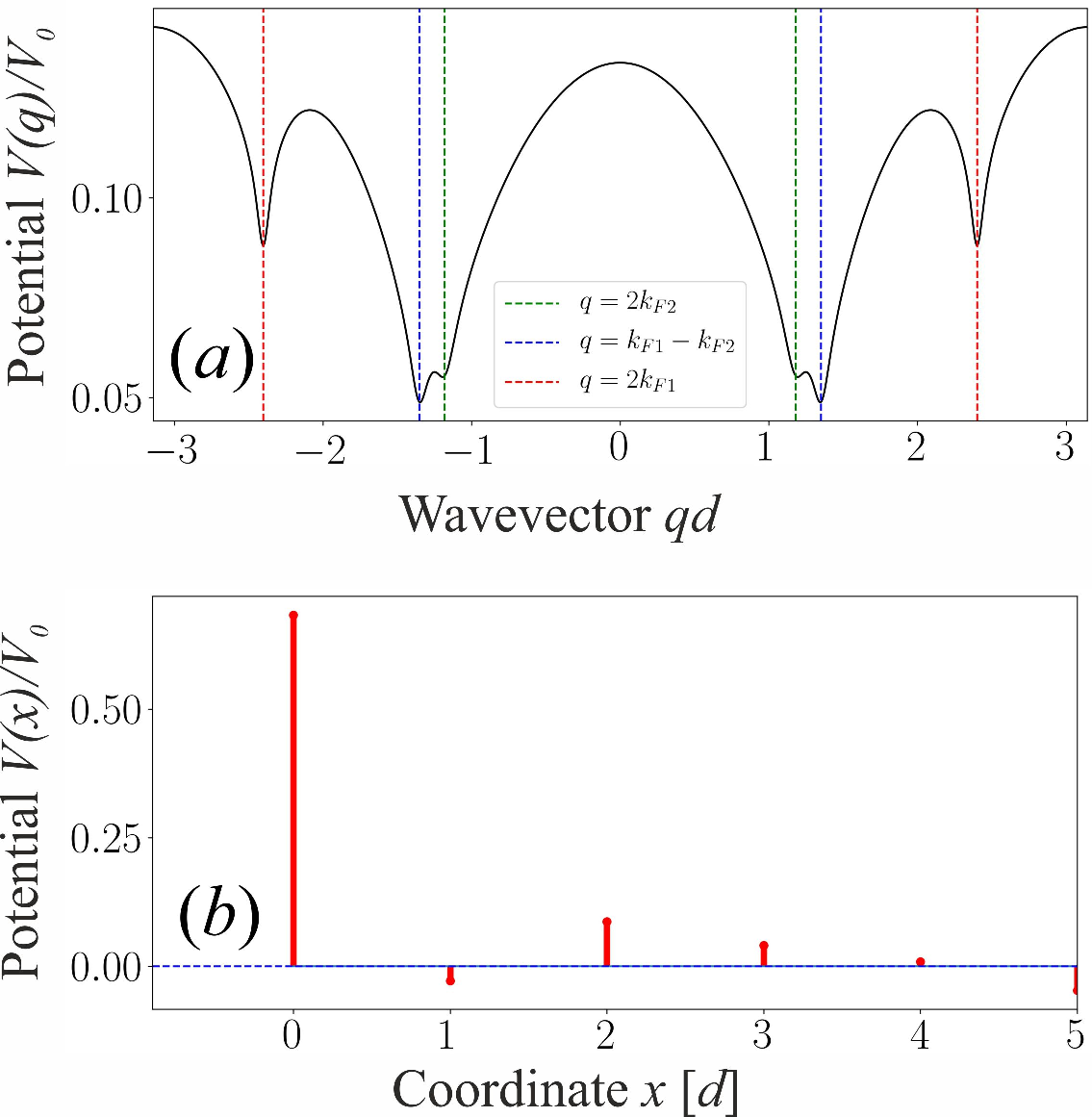}
\caption{The interaction potential for the half-filled system with $t'=0.2t$ under irradiation with $\eta=eEd/\hbar\omega=2.2$ in the reciprocal space (a) and the coordinate space (b).
\label{fig:4}}
\end{figure}
In the coordinate space, this non-monotonic profile projects to the negative value of the interaction potential as shown in Fig.~4b. The attraction at finite distance, which corresponds to these negative values, can favour the Cooper instability with the formation of electron pairs. Following Kohn and Luttinger~\cite{PhysRevLett.15.524}, such a mechanism results in the formation of pairs with finite angular momentum, what leads to superconductivity. Within the mean field approach~\cite{coleman2015introduction}, the order parameter of the superconducting phase, $\Delta(q)$, is defined by the self-consistent equation
\begin{align}
\Delta(k) = -\sum_{k'} V(k-k') \frac{\tanh(\mathcal{E}_{k'}/(2T))\Delta(k')}{2\mathcal{E}_{k'}} \label{eq:self_cons},
\end{align}
where $T$ is the temperature, $\mathcal{E}_k=\sqrt{\Delta^2(k)+(\tilde{\varepsilon}_k-\varepsilon_F)^2}$ is the excitation spectrum, and $\varepsilon_F$ is the Fermi energy. To estimate the critical temperature $T_c$ and the order parameter near this temperature, it should be noted that $|\Delta(k)|\rightarrow 0$ at $T\rightarrow T_c$ and, therefore, one can linearize Eq.~\eqref{eq:self_cons} near the temperature $T_c$ by setting $\mathcal{E}_k\approx |\tilde{\varepsilon}_k-\varepsilon_F|$. Then the linearized  Eq.~\eqref{eq:self_cons} reads
\begin{align}
\Delta(k) = - \sum_{k'} V(k-k') \frac{\tanh(|\tilde{\varepsilon}_k-\varepsilon_F|/2T)}{2|\tilde{\varepsilon}_k-\varepsilon_F|} \Delta(k'). \label{eq:linarized}
\end{align}
It can be seen that the right-hand side of Eq.~\eqref{eq:linarized} acts on the function $\Delta(k)$ as a linear integral operator with the eigenvalues $\lambda=\lambda(T)$. The maximal temperature $T$ which satisfies the self-consistent equation~\eqref{eq:self_cons} is the critical temperature $T_c$, what corresponds to $\lambda(T_c)=1$. As a result of solving this equation, we arrive at the profile of the order parameter slightly below the temperature $T_c$, which is plotted in Fig.~5. It follows from the plots that the order parameter has nodes and is symmetric with respect to the replacement $q\rightarrow -q$. Following the conventional terminology, such a behavior of the order parameter corresponds to the spin-singlet nodal superconductivity. {It should be noted that the appearance of the nodes in the profile of the superconducting gap is typical feature of the Kohn-Luttinger mechanism~\cite{PhysRevLett.15.524}. Within this model, the interaction potential is non-monotonic in the momentum space and changes its sign in the real space. Thus, the potential is repulsive at short distances and attractive at larger distances. Such an interaction potential structure favors the Cooper pair wave function with a complex profile in both the momentum space and the coordinate space. In turn, the profile of the Cooper wave function defines the order parameter profile.}
\begin{figure}[h!]
\centering\includegraphics[width=1.0\columnwidth]{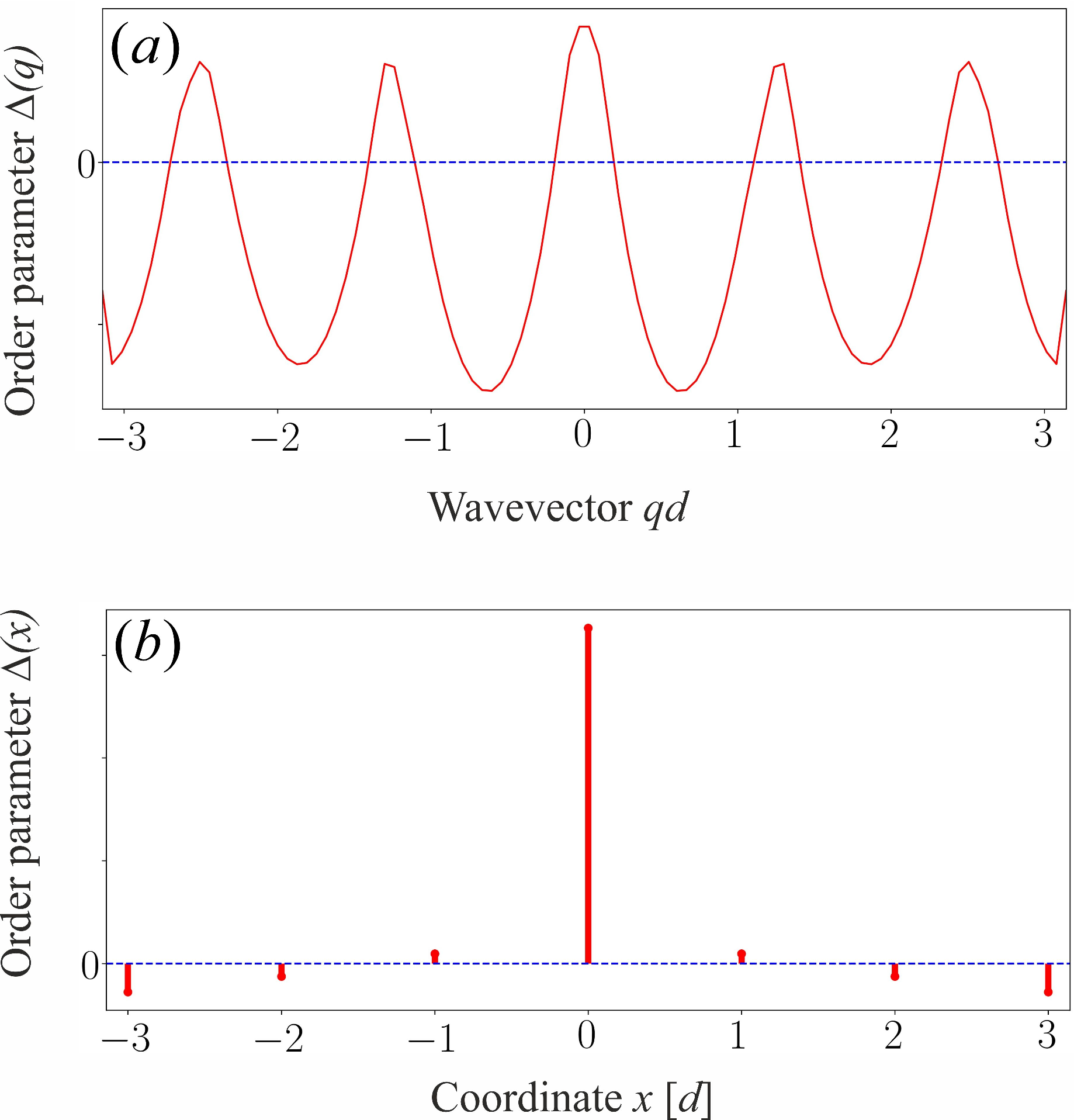}
\caption{Order parameter (in arbitrary units) for the half-filled system with $t'=0.2t$ under irradiation with $\eta=eEd/\hbar\omega=2.2$ in the reciprocal space (a) and the coordinate space (b). \label{fig:5}}
\end{figure}

{Certainly, the RPA approach should be used carefully in the strong coupling regime. Therefore, further analysis of the strong coupling regime may be required for specific materials. However, the superconducting instability emerges at arbitrarily small values of interaction (at least, in the zero temperature limit) provided that the dynamically screened interaction becomes attractive.} It should be noted also that the mean-field results should be treated with care in the one-dimensional system under consideration since fluctuations may play crucial role even in the finite systems. Particularly, the superconducting correlations are $\langle \Delta(x)\Delta(0)\rangle\sim x^{-\gamma}$, where $\gamma$ is the critical exponent. It is well-known that there are multiple  instabilities coexisting with superconducting ones (e.g. ferromagnetic or charge density wave instabilities). In a sufficiently long structure, an instability with the smallest value of $\gamma$, which decays most slowly, is dominant. To find the dominant instability, one has to calculate the critical exponent $\gamma$, what is very complicated problem in the most general case. However, in the particular case of the weak coupling regime, $V_0/\tilde{t}\lesssim 1$, the calculation of $\gamma$ within the Hubbard model can be made with the help of renormalization group technique~\cite{PhysRevB.53.12133} or the Hartree-Fock approximation~\cite{sano2000luttinger}. As a result of the calculation, it has been shown that the superconducting fluctuations are dominant in a wide range of parameters in the phase with the four points at the Fermi surface within the discussed $t-t'$ one-dimensional Hubbard model~\cite{PhysRevB.77.085119}. It should be noted that the strong coupling regime, $V_0/\tilde{t}\gg 1$, can also be realized in our set-up since the dressed hopping coefficient, $\tilde{t}=tJ_0(\eta)$, is very small near a zero of the Bessel function. The discussed one-dimensional Hubbard model with the next-nearest hopping in the strong coupling regime has been analyzed numerically by the density matrix renormalization group methods~\cite{PhysRevB.77.085119}. It follows from this analysis that the spin-singlet superconductive fluctuations remain dominant also in the strong coupling regime if the ratio \eqref{rat} satisfies the condition  $|\tilde{t}'/\tilde{t}|>1$. Since the order parameter behavior plotted in Fig.~5  and discussed above corresponds to the spin-singlet superconductivity, the superconductivity can take place in the considered one-dimensional Hubbard chain of finite length.

It should be noted that the discussed phenomena can also occur in coupled Hubbard chains --- so-called two-leg Hubbard ladders~\cite{Lin_1997} --- which are used, particularly, to describe the wide class of novel unconventional superconductors based on the copper oxides~\cite{matsukawa2004superconductivity}. The two-leg Hubbard ladders comprise two parallel chains of atoms with the intra-chain hopping $t$ and the inter-chain hopping $t'$. In such systems, the ratio \eqref{rat} can be controlled by changing the angle between the field polarization and the system axis, what will result in the same optically induced Lifshitz transition and the corresponding superconductivity. Completing the discussion, it should be stressed that the Lifshitz transitions can be achieved also in a few-layer graphene via the static gate voltage or the electromagnetic dressing~\cite{PhysRevB.96.155432}. Since the emergence of unconventional superconductivity was observed recently in the gating trilayer graphene~\cite{zhou2021superconductivity}, one can expect the experimentally observable similar effects in the systems described by the Hubbard model.

To estimate the experimental feasibility of the effect, one can apply the dressing fields which were recently used to observe the Floquet gap opening in graphene~\cite{mciver2020light}. There the graphene sample was irradiated by short (500 fs) pulses with central frequency $\sim 10$ THz, the mean fluence of $1~\mathrm{mJ cm^{-2}}$ and the peak intensity of $1\mathrm{GW cm^{-2}}$. For these field parameters and the chain period $d\approx 1$nm, the electron-field interaction parameter is $\eta\approx0.7$. Thus, the currently available THz fields are appropriate for observation of the discussed effects.

{It should be noted that the considered system may support many correlated phases, including the Mott insulating phases and the ferromagnetic phases. However, the question of the dominant instability (charge waves or spin density waves, superconducting phase or ferromagnetic phase, etc) is the numerically challenging problem which cannot be solved in a general form. Therefore, this problem should be analyzed carefully for specific structures planned to be studied experimentally.}

\section{Conclusion}
We showed theoretically that an off-resonant high-frequency electromagnetic field can induce the Lifshitz phase transition in periodical structures described by the one-dimensional repulsive Hubbard model with the nearest and next-nearest neighbor hopping. The transition changes the topology of electron energy spectrum at the Fermi level, transforming it from the two Fermi-points to the four Fermi-points. This facilitates the emergence of the spin-singlet superconducting fluctuations in the structure, which can be experimentally observable for the currently available THz fields.

\begin{acknowledgments}
The reported study was funded by the Russian Science Foundation (project 20-12-00001).
\end{acknowledgments}

\appendix
\section{The effective Hamiltonian}
To describe the electron-field interaction with the Hamiltonian \eqref{H}, it is convenient to introduce the two field-dependent dimensionless parameters, $\eta=eEd/\hbar\omega$ and $\xi=t/\hbar\omega$, which can be varied independently by varying the field amplitude $E$ and the field frequency $\omega$. In the present theory, we assume the field frequency to be high enough to satisfy the condition $\xi\ll1$. This allows to describe the dependence of electronic properties on the parameter $\xi$ within the conventional perturbation theory, expanding the Hamiltonian into the $\xi$-power series. As to the field amplitude $E$, it is not assumed to be small. Correspondingly, the parameter $\eta$ can be arbitrary large and will be taken into account accurately.

To proceed, let us expand the periodically time-dependent electron energy $\varepsilon(k-eA(\tau)/\hbar)$ in the Hamiltonian \eqref{H} into the Fourier series
\begin{equation}
\varepsilon(k-eA(\tau)/\hbar)= \tilde{\varepsilon}(k) + \sum_{m} \varepsilon_{m}(k) e^{itm\tau/\hbar\xi},
\end{equation}
where
\begin{equation}
\tilde{\varepsilon}_{m}(k) = -tJ_m(\eta) \cos kd -t'J_m(2\eta) \cos 2kd,
\end{equation}
$J_m(z)$ is the Bessel function of the first kind, $m=\pm1,\pm2,...$ is the order of the Bessel function, and
$\tilde{\varepsilon}(k)=\varepsilon_{0}(k)$. Then the  time-independent part of the Hamiltonian \eqref{H} reduces to the unperturbed Hubbard Hamiltonian with the renormalized tunnelling coefficients, $\tilde{t}=tJ_0(\eta)$ and $\tilde{t}'=t'J_0(2\eta)$. To simplify the following analysis, let us apply the time-dependent unitary transformation $\hat{\cal U}_0=\exp(i\hat{S}_0)$, where
\begin{align}
\hat{S}_0=-i\xi\sum_{k} \sum_m\frac{\varepsilon_{m}(k)}{mt}e^{itm\tau/\hbar\xi}\hat{c}_{k\sigma}^{\dagger}\hat{c}_{k\sigma}
\end{align}
is the transformation operator. Then the transformed Hamiltonian \eqref{H} reads
\begin{align}\label{Hprime}
&\hat{\cal H}'(\tau)=\hat{\cal U}_0\hat{\cal H}\hat{\cal U}_0^{\dagger}-i\hat{\cal U}_0\partial_{\tau}\hat{\cal U}_0^{\dagger}\nonumber\\
&=\sum_k \tilde{\varepsilon}(k) \hat{c}_{k\sigma}^{\dagger}\hat{c}_{k\sigma}+\sum_{k,k',q} \tilde{V}(\tau) \hat{c}_{k-q,\uparrow}^{\dagger}\hat{c}_{k'+q,\downarrow}^{\dagger}\hat{c}_{k'\downarrow}\hat{c}_{k\uparrow},
\end{align}
where
\begin{align}\label{Vt}
&\tilde{V}(\tau)=V_0\exp\left[\xi\sum_m \frac{\varepsilon_{m}(k)+\varepsilon_{m}(k')}{mt}\exp(it m\tau/\hbar\xi)   \right]\nonumber\\
&\times\exp\left[-\xi\sum_m \frac{\varepsilon_{m}(k+q)+\varepsilon_{m}(k-q)}{mt}\exp(it m\tau/\hbar\xi)   \right].
\end{align}
Comparing Eqs.~\eqref{H} and \eqref{Hprime}, one can see that the Hamiltonian \eqref{H} in the field's absence ($A(\tau)=0$) and the Hamiltonian \eqref{Hprime} have the same form with the replacements $\varepsilon(k)\rightarrow\tilde{\varepsilon}(k)$ and $V_0\rightarrow\tilde{V}(\tau)$. Therefore, the function $\tilde{\varepsilon}(k)$ defined by Eq.~\eqref{ekm} should be treated as the electron energy spectrum dressed by the field, whereas the function $\tilde{V}(\tau)$ defined by Eq.~\eqref{Vt} is the dressed interaction potential discussed below.

Since the interaction part of the Hamiltonian \eqref{Hprime} becomes periodically time-dependent, it can be expanded into the Fourier series, $V(\tau)=V_0\sum_l V^{(l)}e^{ilt\tau/\hbar\xi}$, and results for $\xi\ll1$ in $V^{(0)}= V_0+\mathcal{O}(\xi^2)$ and $V^{(j\neq0)} = \mathcal{O}(\xi)$.
Applying the Floquet-Magnus expansion \eqref{Hef} to the Hamiltonian \eqref{Hprime}, one can find that the leading correction to the effective static Hamiltonian is proportional to $\xi \sum_{j\neq 0 }[V^{(j)},V^{(-j)}]\sim \mathcal{O}(\xi^3)$. Thus, up to the linear order in $\xi$, the effective static interaction term remains unchanged, $V(\tau)\approx V_0 + \mathcal{O}(\xi^2)$. As to the non-linear terms, they have a complicated structure describing the field-induced corrections to the interaction between electrons localized at different atoms, including such processes as the electron pair hopping, the exchange interaction,  the dissociation of electron pairs localized at a single atom and the density dependent tunnelling (see, e.g., Ref.~\onlinecite{itin2015effective}). However, the field-induced corrections to the electron interactions decays exponentially as $\xi^{2n}$, where $nd$ is the distance between the interacting electrons ($n=1,2,3,...$). Therefore, these corrections can be neglected as a first approximation under the condition $\xi\ll1$. As a result, we arrive at the effective static Hamiltonian \eqref{H00}.

\end{document}